\documentclass[aps,prbbib,twocolumn,epsf]{revtex4}
\usepackage{amsmath,amssymb}
\usepackage{graphicx} % Include figure files
\usepackage{dcolumn} % Align table columns on decimal point
\usepackage{bm} % bold math

\begin{document}
\draft
\title{\bf Symmetry and transport property of spin current induced spin-Hall effect}
\author{Yanxia Xing$^1$, Qing-feng Sun$^{1,\ast}$,
and Jian Wang$^{2}$ }
\address{
$^1$Beijing National Lab for Condensed Matter Physics and
Institute of Physics, Chinese Academy of Sciences, Beijing 100080,
China\\
$^2$Department of Physics and the Center of Theoretical and
Computational Physics, The University of Hong Kong, Pokfulam Road,
Hong Kong, China }

\begin{abstract}
We study the spin current induced spin-Hall effect that a
longitudinal spin dependent chemical potential $qV_{s=x,y,z}$
induces a transverse spin conductances $G^{ss'}$. A four terminal
system with Rashba and Dresselhaus spin-orbit interaction (SOI) in
the scattering region is considered. By using Landauer-B$\ddot
u$ttiker formula with the aid of the Green function, various spin
current induced spin-Hall conductances $G^{ss'}$ are calculated.
With the charge chemical potential $qV_c$ or spin chemical
potential $qV_{s=x,y,z}$, there are $16$ elements for the
transverse conductances $G^{\mu \nu}_p=J_{p,\mu}/V_{\nu}$ where
$\mu,\nu=x,y,z,c$. Due to the symmetry of our system these
elements are not independent. For the system with $C_2$ symmetry
half of elements are zero, when the center region only exists the
Rashba SOI or Dresselhaus SOI. The numerical results show that of
all the conductance elements, the spin current induced spin-Hall
conductances $G^{ss'}$ are usually much greater (about one or two
orders of magnitude) than the spin Hall conductances $G^{sc}$ and the reciprocal
spin Hall conductances $G^{cs}$. So the spin current induced spin-Hall
effect is dominating in the present device.
\end{abstract}

\pacs{72.25.Mk, 72.20.-i, 73.23.-b}
\maketitle

{\sl Introduction:} Recently, an interesting phenomena, the spin
Hall effect, has been discovered in the spin-orbit interaction
(SOI) system and has attracted considerable attention. In this
effect, a longitudinal external bias (or named charge bias
hereafter) or electric field induces a transverse spin current or
spin accumulations along transverse edges. The spin-Hall effect
can either be extrinsic or intrinsic. The extrinsic spin-Hall
effect is due to the spin dependent scattering\cite{Hirsch} and
has been found a few decades ago. On the other hand, the intrinsic
spin-Hall effect is due to the SOI, predicted by Murakami {\it
et.al.} and Sinova {\sl et.al.} in a Luttinger SOI 3D p-doped
semiconductor\cite{Zhangsc} and a Rashba SOI 2D electron
gas,\cite{Niuq} respectively. Since then, a great deal of sequent
works have focused on this interesting effect. On experimental
side, two groups by Kato {\sl et al.}\cite{experiment1} and
Wunderlich {\sl et al.}\cite{experiment2} have observed the
transverse opposite spin accumulations near two edges of their
devices when the longitudinal voltage bias is added.
More recently, a third group by Valenzuela and Tinkham took the
electronic measurement of the spin Hall effect,\cite{aref1} and
they have observed an induced transverse voltage in a diffusive
metallic conductor when a longitudinal net spin current flows
through it.

Very recently, the reciprocal spin-Hall effect\cite{Onsager,Shen1}
has been investigated, in which a transverse charge conductance is
induced by the driving of a longitudinal spin dependent bias
(named spin bias hereafter). The Onsager reciprocal
relation\cite{Onsager} between the spin-Hall effect and its
reciprocal phenomenon has been found, and the spin-Hall
conductance is predicted to be equal to reciprocal one. So far,
the spin polarization direction considered in most of papers on
the spin-Hall effect\cite{Zhangsc,Niuq} and its reciprocal
effect,\cite{Onsager} is along $z$ direction that is perpendicular
to the 2D electron gas. Since the spin is a vector and it can also
be polarized in the $x$, $y$, as well any $\hat{n}$ directions.
So the direction of the output spin in the two transverse
terminals (i.e. the terminal 2 and 4 in Fig.1) can be in the plane
of the 2D electron gas, e.g. the $x$ or $y$ direction, which has
been studied by Nikoli$\acute{c}$ {\sl et
al}.\cite{Branislav1,Branislav2} The results show that while under
a longitudinal charge bias, the spin currents of the $y$ direction
in the two transverse terminals are equal, i.e. they are
simultaneously flowing out or flowing in (see the left inset in
Fig.1). On the other hand, in the reciprocal spin-Hall effect, the
spin polarization direction for the spin bias in the two
longitudinal terminals (i.e. the terminal 1 and 3 in Fig.1) can
also be in the plane of the 2D electron gas. A few theoretical
works have studied this case,\cite{Shen2} and found that the
charge currents in the terminal 2 and 4 simultaneously flow out or
in (see the left inset in Fig.1), while the spin polarization of
the spin bias is in the $y$ direction. This result is different
from the usual reciprocal spin-Hall effect with the $z$-direction
spin bias, in which the charge current flows in at one transverse
terminal and out at the other one (see the right inset in the
Fig.1).

\begin{figure}%[tbp]
\includegraphics[bb=16mm 15mm 209mm 138mm, width=7cm,totalheight=4.5cm, clip=]{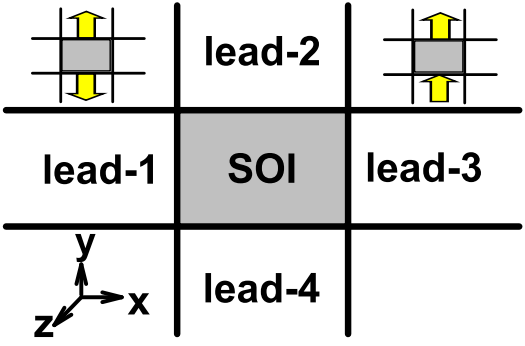}
\caption{ (color online) Schematic diagram for the four terminal
rectangular sample with the Rashba and Dresselhaus SOI in the
center region. The four leads are ideal without SOI. The spin bias
$V_s$ or the charge bias $V_c$ are added on the lead-1 and lead-3,
the induced transverse spin current $J_{p,s}$ or the charge
current $J_{p,c}$ are probed in the lead-2 and lead-4. The left
and right insets depict the directions of the transverse spin or
charge currents when the currents are simultaneously flowing out
($G_2=G_4$) and the currents are flowing in at one terminal and
out at the other one ($G_2=-G_4$), respectively. }
\end{figure}

To explore the vector nature of spin-Hall effect, we set the
longitudinal bias to be spin-dependent and examine the transverse
spin conductance, namely the transverse spin currents are induced
by the longitudinal spin bias, which is named the spin current
induced spin-Hall effect hereafter. So far most of studies focus
on the spin-Hall effect and its reciprocal effect, less attention
has been paid to the spin current induced spin-Hall effect. It is
the purpose of this paper to fill up this gap. We study the spin
current induced spin-Hall effect in a finite mesoscopic system by
using Landauer-B$\ddot{u}$ttiker formula with the aid of the Green
function. Considering a four-terminal ballistic 2D rectangular
region and the center rectangular region having the Rashba and
Dresselhaus SOI (as shown in Fig.1). A spin bias $V_s$ ($s=x$,
$y$, and $z$) is added at the longitudinal terminals 1 and
3.\cite{addnote11} The spin bias can be provided from the device
of the spin cell, which has been suggested by some recent
theoretical works\cite{Sun1} and also been realized in a few
experimental works.\cite{cellex} Under the spin bias $V_s$, the
chemical potentials $\mu$ for the spin-up and spin-down electron
in the terminals 1 and 3 are split, and $\mu_{1,s\uparrow}=
-\mu_{1,s\downarrow} =-\mu_{3,s\uparrow} = \mu_{3,s\downarrow} =
eV/2$, here $s=x$, $y$, and $z$ represents the spin polarization
direction. For comparison, we also consider the charge bias $V_c$
on the longitudinal terminals 1 and 3, in this case
$\mu_{1,s\uparrow}= \mu_{1,s\downarrow} =-\mu_{3,s\uparrow} =
-\mu_{3,s\downarrow} = eV/2$.\cite{note1} The transverse two
terminals 2 and 4 act as the measuring terminals, and their
chemical potentials (or the terminal bias) are set to zero. Due to
the SOI, the spin bias $V_s$ or the charge bias $V_c$ can induced
a transverse spin-Hall currents $J_s$ as well as the charge-Hall
current $J_c$ which will be investigated in this paper. For this
purpose, we use four component vectors ${\bf
J}=(J_{x},J_{y},J_{z},J_c)$ and ${\bf V}=(V_{x},V_{y},V_{z},V_c)$
to represent the spin (charge)-Hall current and spin (charge)
bias, respectively. In the small bias limit, the relationship
between ${\bf J}$ and ${\bf V}$ can be characterized by a $4\times
4$ matrix $G^{\mu\nu}$ ($\mu,\nu \in \{x,y,z,c\}$).

We have studied the relations among these matrix elements from the
symmetry point of view. Four cases have been discussed: (1). only
Rashba SOI is present, (2). only Dresselhaus SOI is present, (3).
both of the Rashba and Dresselhaus SOI are present, (4). the
Rashba and Dresselhaus SOI have the same interaction strength. The
results are summarized in the Table I. All elements are found to
have the property $|G_2^{\mu\nu}| = |G_4^{\mu\nu}|$, where
$G_p^{\mu\nu} = J_{p,\mu}/V_{\nu}$ with $p=2,4$.\cite{note2} This
means that the absolute value $|J_{2,\mu}|$ of the spin ($\mu
=x,y,z$) current or the charge ($\mu=c$) current in the terminal 2
is equal to one $|J_{4,\mu}|$ in the terminal 4 regardless of
longitudinal driving bias $V_{\nu}$. From Table I we see that if
only the Rashba SOI or the Dresselhaus SOI is present in the
center region, half of the matrix elements of the transverse
conductances $G^{\mu\nu}$ are zero. We have also numerically
studied the behavior of these non-zero spin matrix elements
and find that the spin current induced spin-Hall conductances are
much larger (about one or two orders of magnitude) than the
spin Hall conductance as well as its reciprocal one.

{\sl Hamiltonian and Solution:} The system that we considered is
sketched in the Fig.1. The central gray region is the
semiconductor sample in which the SOI of Rashba or/and Dresselhaus
are present. The sample is connected to four ideal non-SOI leads.
The Hamiltonian of the central region is given by $H_0=p^2/2m^*+
V(x,y) + \frac{\alpha}{\hbar}(\sigma_x p_y-\sigma_y
p_x)-\frac{\beta}{\hbar}(\sigma_x p_x-\sigma_y p_y)$, where
$\alpha$ and $\beta$ are the coefficients of the Rashba and
Dresselhaus SOI, $V(x,y)$ is the hard wall confining potential. In
the tight-binding representation, the total Hamiltonian can be
written as:\cite{Hamil,Landau1,addnote12}
\begin{eqnarray}
H&=& \sum\limits_{i} [-t a_{i}^\dagger a_{i+\delta x}
 -t a_{i}^\dagger a_{i+\delta y}
 + i a_{i}^\dagger (V_R \sigma_y +V_D\sigma_x) a_{i+\delta x}\nonumber \\
 &-& i a_{i}^\dagger (V_R\sigma_x +V_D\sigma_y) a_{i+\delta y}]
+H.C
\end{eqnarray}
where $a_i^\dagger=(a_{i \uparrow}^\dagger,a_{i
\downarrow}^\dagger)$, $i$ is the lattice site in the central
region and the leads, $t=\hbar^2/2m^*a^2$ is the hopping matrix
element with the lattice constant $a$, $\delta x$ and $\delta y$
are the unit vectors along the $x$ and $y$ directions, and
$\sigma_x,\sigma_y$ are Pauli matrices. Here $V_R=\alpha/2a$ and
$V_D=\beta/2a$ represent the strength of the Rashba and
Dresselhaus SOI, respectively. $V_R$ and $V_D$ are non-zero only
in the central gray region.

Since there is no SOI in the leads, the particle current
$J_{p,s\sigma}$ in the lead-$p$ ($p$=2,4) with spin at $\pm s$
direction ($\sigma=\uparrow$ or $\downarrow$ stands for the $+s$
or $-s$ direction with $s=x,y,z$) due to the longitudinal spin or
charge bias $V_{\mu}$, can be obtained from the
Landauer-B\"{u}ttiker formula:\cite{Landau}
$J_{p,s\sigma}=(1/h)\sum_{q,\sigma'} T_{ps\sigma,qs'\sigma'}(
\mu_{p,s\sigma} - \mu_{q,s'\sigma'})$ where $\mu_{p,s\sigma}$ and
$\mu_{q,s'\sigma'}$ are the spin-dependent or spin independent
chemical potential related to the bias $V_{\mu}$, which has been
detailed in the the introduction. $T_{qs'\sigma',ps\sigma}$ is the
transmission coefficient from the lead-$p$ with spin $s \sigma$ to
the lead-$q$ with spin $s' \sigma'$.
Note that the spin index $\uparrow$ and $\downarrow$ in the
Hamiltonian (1) represent the spin pointing to the $+z$ and
$-z$-direction, and not pointing to the $\pm x$ and $\pm
y$-direction. So in order to calculate the transmission
coefficient $T_{qs'\sigma',ps\sigma}$ for $s\not=z$ and/or
$s'\not=z$, we need to rotate the z-axis in the spin space in the
lead-$p$ (lead-$q$) to the $s$-direction ($s'$-direction) by
taking an unitary transformation:
\begin{equation}
 \left(\begin{array}{l}
 a'_{i\uparrow}\\a'_{i\downarrow}\end{array} \right) =
 {\bf U}_{s(s')}^{\dagger} \left(\begin{array}{l}
 a_{i\uparrow}\\a_{i\downarrow}\end{array} \right),
\end{equation}
where $i$ is the lattice site in the lead-$p(q)$. The operator
$a_{i\sigma}$ in the center region and other two leads does not
change. The unitary matrix ${\bf U}_{s}$ in the Eq.(2) is
\begin{equation}
 {\bf U}_{s}=\left(
 \begin{array}{cc}
 cos\frac{\theta}{2} & e^{-i\phi} sin\frac{\theta}{2}\\
 e^{i\phi} sin\frac{\theta}{2} & -cos\frac{\theta}{2}
 \end{array}
\right),
\end{equation}
where $(\theta,\phi)$ is the directional angle of $s$-direction
and ${\bf U}_s^{\dagger} \sigma_{s} {\bf U}_{s} =\sigma_z$. Under
this unitary transformation, the Hamiltonian of the leads does not
vary because that the leads' Hamiltonian does not include the
Pauli matrix, only the Hamiltonian that describes the coupling
between the leads and the center region changes. After the unitary
transformation, the z-axis of the spin in the lead-$p(q)$ is in
the $s(s')$-direction, then the transmission coefficient
$T_{qs'\sigma',ps\sigma}$ can be easily obtained as:\cite{Landau}
\begin{equation}
T_{qs'\sigma',ps\sigma}=Tr[{\bf \Gamma}_{q,s'\sigma'}{\bf G}^r
 {\bf \Gamma}_{p,s\sigma} {\bf G}^a] ,
\end{equation}
where the line-width function ${\bf \Gamma}_{p,s\sigma}=i({\bf
\Sigma}_{p,s\sigma}^r-{\bf \Sigma}_{p,s\sigma}^{r\dagger}) ={\bf
\Gamma}_p\otimes {\bf U}_{s} {\bf S}_\sigma {\bf U}^{\dagger}_{s}
$, the retarded self-energy ${\bf \Sigma}_{p,s\sigma}^r = {\bf
\Sigma}^r_p \otimes {\bf U}_{s} {\bf S}_\sigma {\bf
U}^{\dagger}_{s} $ with ${\bf \Gamma}_p$ and ${\bf \Sigma}^r_p$,
respectively, the line-width function and the self-energy of the
lead-$p$ for no-spin-index lattice system. Here the matrix ${\bf
S}_{\sigma}$ is:
\begin{eqnarray}
 {\bf S}_\uparrow =\left(
\begin{array}{cc}
1&0\\
0&0
\end{array}
\right);~~~~~~ {\bf S}_\downarrow =\left(
\begin{array}{cc}
0&0\\
0&1
 \end{array}
 \right)
\end{eqnarray}
The retarded and advanced Green's functions ${\bf G}^{r,a}$ in the
Eq.(4) can be calculated from \cite{Landau} ${\bf G}^r=[{\bf
G}^a]^{\dagger}= \{E_F {\bf I}-{\bf H}_0-\sum_{p,\sigma}{\bf
\Sigma}^r_{p,s\sigma}\}^{-1}$, with the unit matrix ${\bf I}$.
Since $\sum_{\sigma}{\bf \Sigma}^r_{p,s\sigma} ={\bf
\Sigma}_p^r\otimes\sum_{\sigma}({\bf U}_s {\bf S}_{\sigma} {\bf
U}_s^{\dagger})={\bf \Sigma}^r_{p} \otimes {\bf I}$ that is
independent of spin-direction index $s$, the Green functions
${\bf G}^{r,a}$ are also independent of spin-direction index. This
in turn indicates that the Green functions ${\bf G}^{r,a}$ remain
unchanged in the above unitary transformation, in which the spin
axis in the leads is rotated.

Once the particle currents $J_{p,s\sigma}$ ($p=2,4$) are obtained,
the spin current $J_{p,s}$ and the charge current $J_{p,c}$ can be
obtained straightforwardly: $J_{p,c} =
e\{J_{p,s\uparrow}+J_{p,s\downarrow}\}$ and $J_{p,s} =
(\hbar/2)\{J_{p,s\uparrow}-J_{p,s\downarrow}\}$ where $s=x,y,z$.
In fact, it is easy to show that the charge current $J_{p,c}$ is
independent of spin-direction index $s$ of the lead-$p$ (i.e.
$J_{p,x\uparrow}+J_{p,x\downarrow} =
J_{p,y\uparrow}+J_{p,y\downarrow}
=J_{p,z\uparrow}+J_{p,z\downarrow}
=J_{p,\hat{n}\uparrow}+J_{p,\hat{n}\downarrow}$), so the subscript
$s$ is neglected from now on. Then from the current ${\bf
J_p}=(J_{p,x},J_{p,y},J_{p,z},J_{p,c})$ and its driving bias ${\bf
V}=(V_{x},V_{y},V_{z},V_{c})$, we get all the sixteen conductance
matrix elements $G_{p}^{\mu\nu}=J_{p,\mu}/V_{\nu}$ ($\mu,\nu \in
{c,x,y,z}$).\cite{note2} They are the charge Hall conductance
$G^{cc}_p$ describing the transverse charge current $J_{p,c}$
induced by the longitudinal charge bias $V_c$, the spin Hall
conductance $G^{sc}_p$ for the transverse spin current $J_{p,s}$
induced by the longitudinal charge bias $V_c$, the reciprocal spin
Hall conductance $G^{cs}_p$ representing the transverse charge
current $J_{p,c}$ induced by the longitudinal spin bias $V_s$, and
the spin Hall conductance $G^{ss'}_p$ induced by spin current
describing the transverse spin current $J_{p,s}$ induced by the
longitudinal spin bias $V_s$.

{\sl Relations due to symmetry:} We now study the relations among
the sixteen conductance matrix elements $G_{p}^{\mu\nu}$
considering the symmetry of the device. First, our system,
satisfies the time reversal symmetry,\cite{Sun2} so the
transmission coefficients satisfy the relation:
$T_{ps\sigma,qs'\sigma'} = T_{qs'{\bar \sigma'},ps{\bar \sigma}}$,
which determines the properties of the transverse conductances.
Second, if the shape of the device has the geometrical symmetry,
the results can be greatly simplified and many matrix elements are
zero. In the following, we consider the rectangular center region
that has $C_2$ symmetry.

For the rectangular sample, the shape of the device [i.e the
confining potential $V(x,y)$] is invariant under the rotation
transformation $C_2$ by rotating an angle $\pi$ around the
$x,y,z$-axis at the center point. But the SOI part of the
Hamiltonian is varied under the space rotation transformation
$C_2$. In order to keep the invariance of the total Hamiltonian
$H$, we construct the unitary transformation $U=U_0\otimes U_s$,
where $U_0$ and $U_s$ are the rotation transformation in the real
space and in the spin space, respectively. In the following, we
list all of the unitary transformations $U$, under which the total
Hamiltonian $H$ is invariance (i.e. $U^{\dagger} H U= H$). We then
derive the relations among the transverse conductance matrix
elements from the symmetry. We consider four cases of SOI: (1).
with Rashba SOI only, (2). with Dresselhaus SOI only, (3). with
both the Rashba and Dresselhaus SOI, (4). similar to (3) but the
strengths of the Rashba and Dresselhaus SOI are equal. To
illustrate the derivation, we shall discuss the case of the Rashba
SOI case in detail. For other three cases, similar discussions
apply and we only give the results.

First we study the case with only Rashba SOI using the following
transformations.

(i) $U_1=C_{2z}\otimes exp(-i\frac{\pi}{2}\sigma_z)$, which
performs the transformation: $x\rightarrow -x$, $y\rightarrow -y$,
$z\rightarrow z$, and $\sigma_x\rightarrow -\sigma_x$,
$\sigma_y\rightarrow -\sigma_y$, $\sigma_z\rightarrow \sigma_z$.
Under this transformation $U_1$, the system (including both of the
real space and the spin space) is rotated by an angle $\pi$ around
the $z$-axis, which leads $J_{2,x\sigma}
\rightarrow J_{4,x\bar{\sigma}}$ and $\mu_{1,x\sigma} \rightarrow
\mu_{3,x\bar{\sigma}}$, so the spin current
$J_{2,x}=(\hbar/2)\{J_{2,x\uparrow}-J_{2,x\downarrow}\}\rightarrow
(\hbar/2)\{J_{4,x\downarrow}-J_{4,x\uparrow}\} =-J_{4,x}$, and the
spin chemical potential $V_x$ remains unchanged because the charge
chemical potential $V_c$ and the $\sigma_x$ change signs due to
the rotations in real space and spin space, respectively.
Similarly, the transformation for the others $J_{2,\mu}$ and
$V_{\mu}$ can be also deduced as:
\begin{equation}\label{symme1}
\begin{array}{lll}
 J_{2,c} & \rightarrow & J_{4,c} \\
 J_{2,x} & \rightarrow  & -J_{4,x} \\
 J_{2,y} & \rightarrow  & -J_{4,y}\\
 J_{2,z} & \rightarrow & J_{4,z}
\end{array},\hspace{10mm}
\begin{array}{lll}
V_{c}& \rightarrow &-V_{c} \\
V_{x}& \rightarrow &V_{x} \\
V_{y}& \rightarrow &V_{y} \\
V_{z}& \rightarrow &-V_{z}
\end{array},
\end{equation}
Then the relations among the matrix elements of the transverse
conductance can be obtained straightforwardly. For example,
$G_2^{xx} =J_{2,x}/V_{x} = -J_{4,x}/V_{x} = -G_{4}^{xx}$, etc.

(ii) $U_2=C_{2y}\otimes exp(-i\frac{\pi}{2}\sigma_x)$, which
performs the transformation: $x\rightarrow -x$, $y\rightarrow y$,
$z\rightarrow -z$, and $\sigma_x\rightarrow \sigma_x$,
$\sigma_y\rightarrow -\sigma_y$, $\sigma_z\rightarrow -\sigma_z$.
This transformation $U_2$ is equivalent to rotate the real space
by $\pi$ around the $y$-axis and to rotate the spin by $\pi$
around the $x$-axis. Under the unitary transformation $U_2$,
$J_{p,\mu}$ ($p=2,4$) and $V_{\mu}$ are transformed into:
\begin{equation}\label{symme2}
\begin{array}{lll}
 J_{2(4),c} & \rightarrow & J_{2(4),c} \\
 J_{2(4),x} & \rightarrow  & J_{2(4),x} \\
 J_{2(4),y} & \rightarrow  & -J_{2(4),y}\\
 J_{2(4),z} & \rightarrow & -J_{2(4),z}
\end{array},\hspace{10mm}
\begin{array}{lll}
V_{c}& \rightarrow &-V_{c} \\
V_{x}& \rightarrow &-V_{x} \\
V_{y}& \rightarrow &V_{y} \\
V_{z}& \rightarrow &V_{z}
\end{array},
\end{equation}
Then $G_{p}^{cc} =J_{p,c}/V_{c} = J_{p,c}/(-V_{c}) = -G_{p}^{cc}$,
so $G_{p}^{cc} =0$ ($p=2,4$). In fact, from $U_2^{\dagger} H U_2
=H$, we obtain that eight matrix elements of the transverse
conductance are zero, $G_{p}^{cc} = G_{p}^{cx}= G_{p}^{xc}=
G_{p}^{xx}= G_{p}^{yy}= G_{p}^{yz}= G_{p}^{zy}= G_{p}^{zz}=0$
($p=2,4$).

(iii) $U_3=C_{2x}\otimes exp(-i\frac{\pi}{2}\sigma_y)$, which
performs the transformation: $x\rightarrow x$, $y\rightarrow -y$,
$z\rightarrow -z$, and $\sigma_x\rightarrow -\sigma_x$,
$\sigma_y\rightarrow \sigma_y$, $\sigma_z\rightarrow -\sigma_z$.
This transformation $U_3$ rotates the real space by $\pi$ around
the $x$-axis and the spin by $\pi$ around the $y$-axis. Under this
transformation, $J_{p,\mu}$ ($p=2,4$) and $V_{\mu}$ are
transformed into:
\begin{equation}\label{symme3}
\begin{array}{lll}
 J_{2,c} & \rightarrow & J_{4,c} \\
 J_{2,x} & \rightarrow  & -J_{4,x} \\
 J_{2,y} & \rightarrow  & J_{4,y}\\
 J_{2,z} & \rightarrow & -J_{4,z}
\end{array},\hspace{10mm}
\begin{array}{lll}
V_{c}& \rightarrow &V_{c} \\
V_{x}& \rightarrow &-V_{x} \\
V_{y}& \rightarrow &V_{y} \\
V_{z}& \rightarrow &-V_{z}
\end{array},
\end{equation}
It is worth mentioning that the above three unitary
transformations are not independent on each other, and we have
$U_3 U_1 =U_2$. So Eq.(\ref{symme2}) can be obtained from
Eq.(\ref{symme1}) and Eq.(\ref{symme3}).

Combining Eqs.(\ref{symme1},\ref{symme2},\ref{symme3}), the
relations among the matrix elements of the transverse conductances
$G^{\mu\nu}_p$ are obtained and summarized in the Table I(a). The
sixteen conductance matrix elements are arranged in the matrix
form, in which the columns (labelled by $J_{\mu}$) denote the
measured transverse conductances (from lead 2 or 4) and the rows
are the longitudinal bias $V_{\nu}$ in lead 1 or 3. The zero
conductance matrix elements are indicated by the symbol '0'. The
symbol '$+$ ($-$)' denotes the non-zero conductance matrix
elements $G_p^{\mu\nu}$, which have the same (opposite) signs for
the lead-2 and lead-4, i.e. $G_2^{\mu\nu}=\pm G_4^{\mu\nu}$. For
some pair of conductance matrix elements, they may have the same
value or differ by a sign, e.g. $G_{p}^{cz}=G_{p}^{zc}$, $p=2,4$,
which is marked by the letter symbols in the Table I. Of which,
different symbols present different conductance values, and the
symbol $a$ (or $\bar{a}$) are for $G_2^{\mu\nu}=a$ (or
$G_2^{\mu\nu}=-a$).

\begin{table}
\caption[Symmetry]{Symmetry of the transverse spin or charge
conductance for the system with the Rashba and/or Dresselhaus SOI:
(a), $V_R\neq0$, $V_D=0$; (b), $V_R=0$, $V_D\neq0$; (c), $V_R\neq
V_D\neq0$ and (d), $V_R=V_D\neq0$. The symbol '0' indicates the
corresponding $G_{2(4)}^{\mu\nu}=0$, and the symbol '$+$ ($-$)'
denotes $G_2^{\mu\nu}=\pm G_4^{\mu\nu}$. }
\begin{tabular}{|c|ccccc||c|ccccc|}
\hline
{\rule[-1mm]{0mm}{5.0mm}}
(a)& $V_{x}$ & $V_{y}$ & $V_{z}$&
$\genfrac{}{}{0pt}{}{\shortmid}{\shortmid}$
& $V_{c}$&
(b)& $V_{x}$ & $V_{y}$ & $V_{z}$ &
$\genfrac{}{}{0pt}{}{\shortmid}{\shortmid}$
&$V_{c}$\\
\hline
 {\rule[0mm]{0mm}{4mm}}$J_{x}$ & $0$ & $-$ & $+$ &
$\genfrac{}{}{0pt}{}{\shortmid}{\shortmid}$
&$0$ &
 $J_{x}$ & $0$ & $-$ & $0$ &
$\genfrac{}{}{0pt}{}{\shortmid}{\shortmid}$
&$+$ \\
 {\rule[0mm]{0mm}{3.5mm}}$J_{y}$ & $-$ & $0$ & $0$ &
$\genfrac{}{}{0pt}{}{\shortmid}{\shortmid}$
&$+$ &
 $J_{y}$ & $-$ & $0$ & $+$ &
$\genfrac{}{}{0pt}{}{\shortmid}{\shortmid}$
&$0$ \\
 {\rule[0mm]{0mm}{3.5mm}}$J_{z}$ & $+$ & $0$ & $0$ &
$\genfrac{}{}{0pt}{}{\shortmid}{\shortmid}$
&$-_a$ &
 $J_{z}$ & $0$ & $+$ & $0$ &
$\genfrac{}{}{0pt}{}{\shortmid}{\shortmid}$
&$-_c$ \\
-~-~- &-~-~- &-~-~- &-~-~- &$\genfrac{}{}{0.1pt}{}{\shortmid}{\shortmid}$ &-~-~- &
-~-~- &-~-~- &-~-~- &-~-~- &$\genfrac{}{}{0.1pt}{}{\shortmid}{\shortmid}$ &-~-~- \\
 {\rule[-2mm]{0mm}{3.5mm}}$J_{c}$  & $0$ & $+$ & $-_a$ &
$\genfrac{}{}{0pt}{}{\shortmid}{\shortmid}$
&$0$ &
 $J_{c}$  & $+$ & $0$ & $-_c$ &
$\genfrac{}{}{0pt}{}{\shortmid}{\shortmid}$
&$0$ \\
\hline\hline
(c)& {\rule[-1mm]{0mm}{5.0mm}}$V_{x}$ & $V_{y}$ & $V_{z}$ &
$\genfrac{}{}{0pt}{}{\shortmid}{\shortmid}$
&$V_{c}$&
(d)& $V_{x}$ & $V_{y}$ & $V_{z}$ &
$\genfrac{}{}{0pt}{}{\shortmid}{\shortmid}$
&$V_{c}$\\
\hline
 {\rule[0mm]{0mm}{4mm}}$J_{x}$ & $-$ & $-$ & $+$ &
$\genfrac{}{}{0pt}{}{\shortmid}{\shortmid}$
&$+$ &
 $J_{x}$ & $-_r$ & $-_s$ & $+_o$ &
$\genfrac{}{}{0pt}{}{\shortmid}{\shortmid}$
&$+_e$ \\
 {\rule[0mm]{0mm}{3.5mm}}$J_{y}$ & $-$ & $-$ & $+$ &
$\genfrac{}{}{0pt}{}{\shortmid}{\shortmid}$
&$+$ &
 $J_{y}$ & $-_s$ & $-_r$ & $+_{\bar{o}}$ &
$\genfrac{}{}{0pt}{}{\shortmid}{\shortmid}$
&$+_e$ \\
 {\rule[0mm]{0mm}{3.5mm}}$J_{z}$ & $+$ & $+$ & $-$ &
 $\genfrac{}{}{0pt}{}{\shortmid}{\shortmid}$ &$-$ & $J_{z}$ &
$+_u$ & $+_u$ & $-$ & $\genfrac{}{}{0pt}{}{\shortmid}{\shortmid}$
&$0$ \\
-~-~- &-~-~- &-~-~- &-~-~- &$\genfrac{}{}{0.1pt}{}{\shortmid}{\shortmid}$ &-~-~- &
-~-~- &-~-~- &-~-~- &-~-~- &$\genfrac{}{}{0.1pt}{}{\shortmid}{\shortmid}$ &-~-~- \\
 {\rule[-2mm]{0mm}{3.5mm}}$J_{c}$  & $+$ & $+$ & $-$ &
 $\genfrac{}{}{0pt}{}{\shortmid}{\shortmid}$ &$-$ & $J_{c}$  &
$+_v$ & $+_{\bar{v}}$ & $0$ &
$\genfrac{}{}{0pt}{}{\shortmid}{\shortmid}$
&$-$ \\
\hline
\end{tabular}
\end{table}

To examine the system with Dresselhaus SOI only, we need only
change $\sigma_x$ to $\sigma_y$ and $\sigma_y$ to $\sigma_x$ in
the above Hamiltonian with Rashba SOI and unitary transformations.
As a result, the above three transformations are changed to
$U_1=C_{2z}\otimes exp(-i\frac{\pi}{2}\sigma_z)$,
$U_2=C_{2y}\otimes exp(-i\frac{\pi}{2}\sigma_y)$ and
$U_3=C_{2x}\otimes exp(-i\frac{\pi}{2}\sigma_x)$. Under these
transformations, we find all the symmetry relations among the
sixteen conductances and the results are listed in the Table I(b).
For the case of having both the Rashba and Dresselhaus SOI
($V_R\not=0$ and $V_D\not=0$), only the transformation $U_1$ can
keep the Hamiltonian invariant, and it leads the symmetry
relations shown in the Table I(c). For the case of
$V_R=V_D\not=0$, except $U_1$, there is an additional symmetry
$U_4$: ${\bf I} \otimes (\sigma_{x(y)}\rightarrow \sigma_{y(x)},
\sigma_{z}\rightarrow \sigma_{z})$, which leads to the symmetry
between the $s_x$ and $s_y$, and the relations of the conductance
elements $G_p^{\mu\nu}$ are expressed in the Table I(d). In fact,
$U_4$ only takes the transformation in the spin space, so the
symmetry $U_4$ always exists and $G_p^{zc}=G_p^{cz}=0$ for any
confining potential $V(x,y)$ (not requiring the rectangular
sample).

From these symmetry relations, we conclude: (i) All the
conductance matrix elements are found to obey the relation
$|G_2^{\mu\nu}| = |G_4^{\mu\nu}|$, i.e.,  the absolute value of
$|J_{2,\mu}|$ of the spin ($\mu =x,y,z$) or the charge ($\mu=c$)
current in the terminal 2 is equal to the one of $|J_{4,\mu}|$ in
the terminal 4. Furthermore, $G_2^{\mu\nu} =- G_4^{\mu\nu}$ for
the eight block-diagonal elements in Table I, and $G_2^{\mu\nu} =
G_4^{\mu\nu}$ for the eight non-block-diagonal elements. (ii)
Exchanging $\sigma_x$ and $\sigma_y$, the system with Rashba SOI
is same as that with Dresselhaus SOI, which can be seen from
the Table I(a) and I(b). (iii) If only the Rashba SOI or
Dresselhaus SOI is present in the center region, half of matrix
elements $G^{\mu\nu}$ are zero.\cite{note3} (vi) For the usual
spin Hall effect or its reciprocal effect (in which the spin is
polarized along the $z$-direction), the transverse spin current or
charge current are conserved (i.e. $G^{zc(cz)}_2 =
-G^{zc(cz)}_4$). Furthermore, $G^{zc}$ and $G^{cz}$ satisfy the
Onsager relation:
$G_{2(4)}^{zc}=G_{2(4)}^{cz}$.\cite{Onsager,note4} While for
$G^{cy}$ and $G^{yc}$, as well the non-block-diagonal elements in
Table I, they either flow in or out of the terminals 2 and 4 (i.e.
$G_2=G_4$), and do not satisfy the Onsager relation, because that
the Onsager relation requires that the currents in the two
transverse terminals 2 and 4 must be $J_2=-J_4$.

{\sl Numerical results and Discussion}: In the following, we
numerically study the non-zero conductance matrix elements
$G_p^{\mu\nu}$, and we mainly focus the spin Hall conductance
induced by the spin current $G_p^{ss'}$ ($s,s' \in x,y,z$). For
simplicity, only the case of Rashba SOI is shown, and the results
for the other three case have similar physics. In the numerical
calculation, we consider the square scattering center region for
convenience. But notice that the symmetry relation in Table I is
valid for the rectangular scattering center region. In the
calculation the electron effective mass $m^*$ is set $0.05
m_e$,\cite{Gui,note5} and the Fermi energy $E_F=-3.8 t$ (in
Fig.4,5) which is near the band bottom $-4t$, with $t=1$ as a
energy unit and the corresponding lattice constant $a \approx
2.6nm$.\cite{Branislav1,Branislav2} The size of center region $L$
is chosen in the same order as the spin precession length $L_{SO}$
($L_{SO} \equiv \pi a t/2V_R$) over the precessing angle $\pi$. If
taking $V_R =0.03t$ (corresponding to $\alpha\approx 1\times
10^{-11}$ eV$\cdot$m), $L_{SO}\approx 50 a$.

\begin{figure}%[tbp]
\includegraphics[bb=12mm 9mm 202mm 196mm, width=8.0cm,totalheight=8.5cm,clip=]{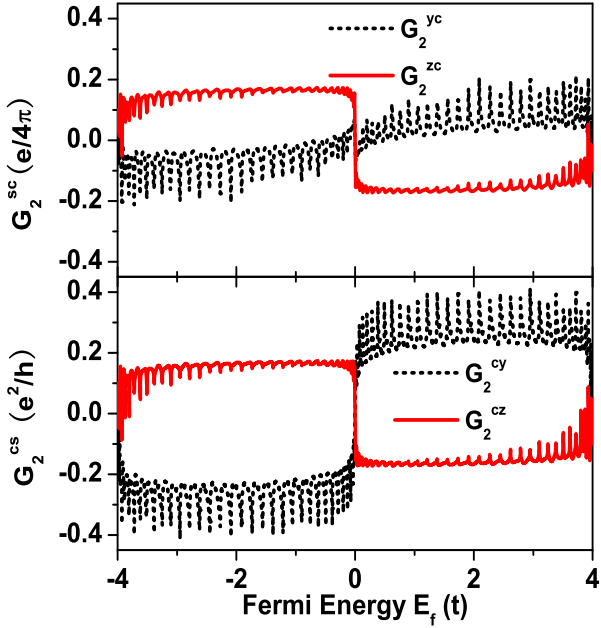}
\caption{(Color online) The spin Hall conductances $G^{sc}$ and
the reciprocal spin Hall conductances $G^{cs}$ {\sl vs} Fermi
energy $E_f$, with the parameters $L=34a$ and $V_R=0.03t$.}
\end{figure}

\begin{figure}%[tbp]
\includegraphics[bb=11mm 10mm 181mm 118mm, width=8.0cm,totalheight=5cm,clip=]{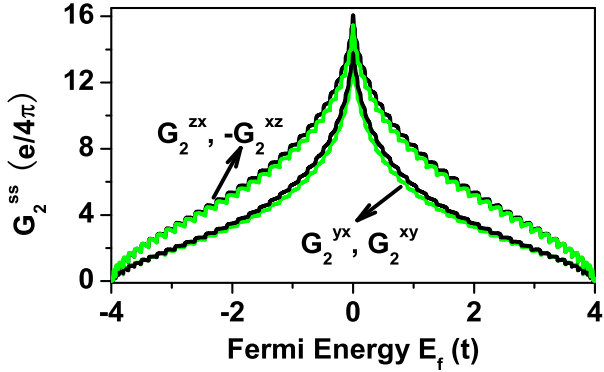}
\caption{(Color online) The non-zero spin current induced spin
Hall conductance elements $G_2^{xy}$ (black), $G_2^{yx}$ (green or
gray), $G_2^{zx}$ (black), and $-G_2^{xz}$ (green or gray) {\sl
vs} Fermi energy $E_f$. The parameters are: $L=34a$ and
$V_R=0.03t$. }
\end{figure}

First, Fig.2 shows the spin Hall conductance $G^{zc}$, the
reciprocal spin Hall conductance $G^{cz}$, the in-plan spin Hall
conductance $G^{yc}$, and its reciprocal conductance $G^{cy}$. We
see that they are odd functions of the Fermi energy $E_f$ and
vanish at the half-filled band with $E_f =0$. The behaviors of the
spin Hall conductance $G^{zc}$ and $G^{yc}$ are similar to the
previous works done by Nikoli$\acute{c}$
et.al.\cite{Branislav1,Branislav2} From the Fig.2, we can find
that out-plane matrix elements $G^{zc}=G^{cz}$, which satisfies
the Onsager relation.\cite{Onsager} On the other hand, the
in-plane matrix elements $G^{yc}$ and $G^{cy}$ do not obey the
Onsager relation, because $G^{yc}$ and $G^{cy}$ are simultaneously
flow out of or into the two transverse terminals 2 and 4 (as shown
in the left inset of Fig.1), and they do not satisfy the
corresponding condition. $G^{yc}$ is smaller than $G^{cy}$, and
both of them strongly depend on the Fermi energy.

Second, we examine the spin Hall conductances induced by spin
current $G^{ss'}$ ($s,s' \in x,y,z$). Fig.3 shows the four
non-zero matrix elements $G^{ss'}$ versus the Fermi energy $E_f$.
Different from $G^{sc}$ and $G^{cs}$, the spin current induced
spin Hall conductances $G^{ss'}$ are even functions of $E_f$, and
$|G^{ss'}|$ reaches the largest value at the half-filled band with
$E_f =0$. In particular, the spin current induced spin Hall
conductances $G^{ss'}$ are much larger (at least one order of
magnitude larger) than $G^{sc}$ or $G^{cs}$ (see Fig.2 and Fig.3).
This means that the spin current induced spin Hall effect is the
dominating effect in the SOI system. For the spin current induced
spin Hall conductances $G^{ss'}$, they do not obey the Onsager
relations, in general $G^{xy}\not= G^{yx}$ and $G^{xz}\not=
G^{zx}$. This is because that the present device is not simply
four-terminal system, more like the eight-terminal system with the
index of both the terminal and the spin. Also because that for the
circuit of the Onsager relations, it requires that the boundary
condition of the transverse terminals 2 and 4 is $J_2 =-J_4$, i.e.
in the external circuit, the terminals 2 and 4, and the terminals
1 and 3, are connected, respectively. However, in the present
system, the current $G_2^{\mu\nu}\not=-G_4^{\mu\nu}$ for the
non-block-diagonal elements. On the other hand, for the small
$V_R$ (e.g. $V_R<0.08t$), $|G^{xy}|$ and $|G^{xz}|$ are still
approximatively equal to $|G^{yx}|$ and $|G^{zx}|$, respectively.

The spin current induced spin Hall conductances $G^{ss'}$ versus
the sample size $L$ and the Rashba SOI strength $V_R$ are plotted
in Fig.4 and Fig.5, respectively. With the increasing of the size
of center region $L$, the conductances $G^{yx}$ and $G^{xy}$ are
greatly enhanced (see Fig.4), because that the transverse
terminals 2 and 4 are much wider at the large $L$. But at a few
special sizes $L$ (e.g. $L=26, 46$ for $V_R=0.06t$), $G^{xy}$ and
$G^{yx}$ are anomalous, in which $G^{xy}$ and $G^{yx}$ are
minimum. This is because that a subband passes the Fermi energy
$E_f$. The conductances $G^{zx}$ and $G^{xz}$ have the similar
behaviors with $G^{yx}$ and $G^{xy}$ while in the small $V_R$
region, i.e. the corresponding spin precession length $L_{SO} =\pi
a t/2V_R > L$. But for the large $V_R$ case (i.e. $L_{SO} <L$),
$G^{zx}$ and $G^{xz}$ are not regular with the size $L$. Because
in the present spin Hall device, the quantum states are
extended.\cite{Xing}

On the other hand, with increasing of the Rashba SOI strength
$V_R$, the conductances $G^{ss'}$ are oscillatory (see Fig.5). For
the larger $V_R$, $G^{ss'}$ shows a chaotic behavior. Furthermore,
at the large $V_R$, $|G^{xy}|$ and $|G^{xz}|$ are not equal to
$|G^{yx}|$ and $|G^{zx}|$, i.e. the Onsager relations for the spin
current induced spin Hall conductances $G^{ss'}$ are completely
violated.

\begin{figure}%[tbp]
\includegraphics[bb=11mm 10mm 198mm 289mm, width=6.5cm,totalheight=9.0cm,clip=]{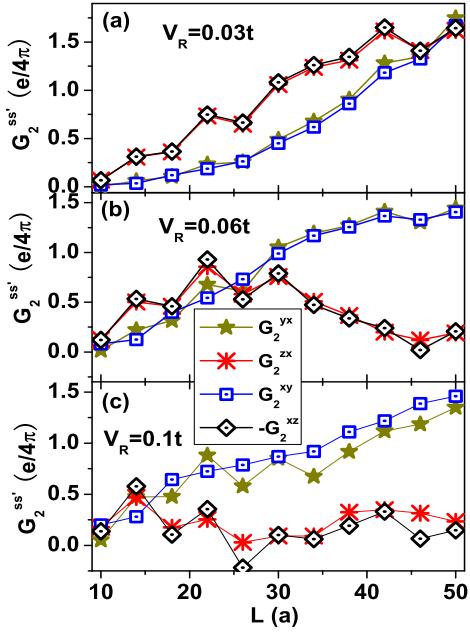}
\caption{(Color online) The spin current induced spin Hall
conductances $G^{ss'}$ {\sl vs} sample size $L$ for different SOI
strength $V_R=0.03$ (a), $V_R=0.06$ (b), and $V_R=0.1$ (c). The
curves of $G^{xy}_{2}$ and $G^{yx}_{2}$, $-G^{xz}_2$ and
$G^{zx}_{2}$ are nearly coincident for the small $V_R=0.03$ and
$0.06$. The Fermi energy is: $E_f=-3.8t$.}
\end{figure}

\begin{figure}%[tbp]
\includegraphics[bb=11mm 10mm 192mm 162mm, width=7.0cm,totalheight=5.2cm,clip=]{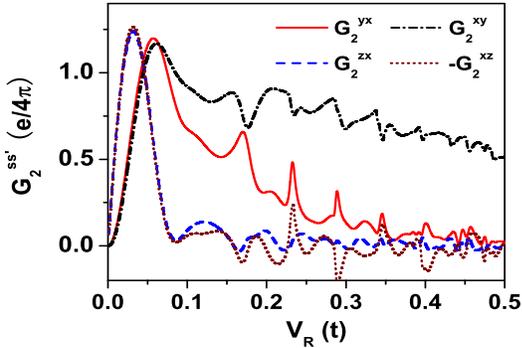}
\caption{(Color online) The spin current induced spin Hall
conductances $G^{ss'}$ {\sl vs} the Rashba SOI strength $V_R$, with
the parameters $L=34a$ and $E_f=-3.8t$.}
\end{figure}

At last, we investigate how the spin current induced spin-Hall
conductances $G^{ss'}$ are affected by the disorders. In the
previous calculation, the on-site energies $\epsilon_i$ (i.e. in
the term $\sum_i \epsilon_i a_i^{\dagger} a_i$) in Eq.(1) are taken
to be zero when there is no disorders. To consider the
effect of disorder, random on-site potentials $\epsilon_i$
in the center region are added with an uniform
distribution $[-W/2, W/2]$ with disorder strength $W$. The
conductance is obtained by averaging over up to 1000 disorder
configurations. Fig.6 shows the normalized conductances
$G^{ss'}_{ratio}$ [$G^{ss'}_{ratio} \equiv G^{ss'}(W\neq
0)/G^{ss'}(W=0)$] versus the disorder strength $W$ for the
different size $L$. From the Fig.6, we can see that the spin
current induced spin Hall conductances $G^{ss'}$ decrease with the
increasing of the disorder strength $W$, but $G^{ss'}$ keep the
large value while $W<t$. This behavior is similar to the spin
Hall conductances which have been investigated
recently.\cite{Sheng} Hence, in the dirty case, the spin current
induced spin Hall effect is still dominant in the finite 2D SOI
system. In additiona, for the different size of the device, the
relation of $G^{xy}_{ratio}$ and $G^{xz}_{ratio}$ versus $W$
almost remain the same. On the other hand, $G^{yx}_{ratio}$ and
$G^{zx}_{ratio}$ can keep larger value in the system with small
size than the system with big size for fixed disorder strength
$W$.

\begin{figure}%[tbp]
\includegraphics[bb=9mm 9mm 201mm 148mm, width=8cm,totalheight=6.5cm,clip=]{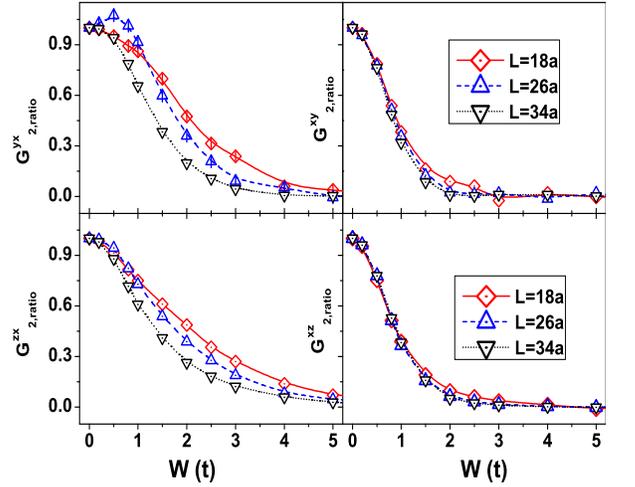}
\caption{(Color online) the normalized conductances
$G^{ss'}_{ratio}$ [$G^{ss'}_{ratio} =G^{ss'}(W\neq
0)/G^{ss'}(W=0)]$ {\sl vs} the disorder strength $W$ with the
different size $L=18a$, $26a$, and $34a$. The other parameters are
$E_f=-3.8t$ and $V_R=0.03$.}
\end{figure}

{\sl Summary}: The spin current induced spin Hall effect is
investigated in a four terminal system with the center region
having the spin-orbit interaction (SOI). Because of the vector
nature of spin, the charge and three spin components form a
$4\times 4=16$ transverse conductance matrix whose matrix elements
include the spin current induced spin Hall conductances $G^{ss'}$
($s,s'\in x,y,z$), the spin Hall conductances $G^{sc}$, the
reciprocal spin Hall conductances $G^{cs}$, and the charge Hall
conductance $G^{cc}$. Of these matrix elements, we found that in
general, $G^{ss'}$ are much larger (about one or two orders) than
the others. This means that the spin current induced spin Hall
effect is the dominating effect in the present device. By
analyzing the system's symmetry, the relations among these
conductance matrix elements are found. The results indicated that
eight matrix elements ($G^{xx/xy/yx/yy}$ and $G^{zz/zc/cz/cc}$)
have conserved quantities as the usual Hall effect. But the other
eight matrix elements correspond to the current simultaneously
flowing out of or into two transverse terminals (as shown in the
left inset of Fig.1), which is different from the usual Hall
effect. When only the Rashba SOI or Dresselhaus SOI is present in
the device, half of matrix elements are found to be zero.

{\bf Acknowledgments:} This work was supported by the Knowledge
Innovation Project of the Chinese Academy of Sciences and
NSF-China under Grant No. 90303016, No. 10474125, and No.
10525418. J. W. is supported by RGC grant (HKU 7044/05P)
from the government SAR of Hong Kong.


\begin{thebibliography}{00}
\bibitem{addnote11}
In our model, a pure spin bias $V_s$ (with zero charge bias $V_c$)
is added at the longitudinal terminals. Suppose the
longitudinal terminals 1 and 3 are replaced by the ferromagnetic leads
and we apply both the spin bias and the charge bias. In this
case, the spin and charge currents in the transverse terminals 2
and 4 can easily be obtained from $J_{\mu}=G^{\mu\nu}V_{\nu}$ with
the aid of $G^{\mu\nu}$. But now the spin current induced
spin-Hall effect, the spin Hall effect, the reciprocal spin Hall
effect, and the Hall effect, will mix together.
\bibitem{note1}
Under the charge bias $V_c$, the chemical potentials $\mu_{1/3}$
are spin indenpendent. So the results (e.g. the charge or spin
current) are completely the same, regardless of the spin index $s$
used in $\mu_{1,s\uparrow}= \mu_{1,s\downarrow}
=-\mu_{3,s\uparrow} = -\mu_{3,s\downarrow} = eV/2$.
\bibitem{addnote12}
Here the four leads are also described by the discrete
tight-binding Hamiltonian, which is the same as the
Ref.\cite{Hamil}.
\bibitem{note3}
In Ref.\cite{Branislav2} the matrix element $G^{xc}$ is very
small, but it is not zero. This is different from the our present
results [$G^{xc}=0$ exactly, see Table I(a)] although the system
is completely the same. They may have neglected a factor of
$G^{out}_{22}-G^{in}_{22}$ in their equation (8).
\bibitem{note4}
Notice that $G_{2(4)}^{zc}=G_{2(4)}^{cz}$ is valid only for the
square scattering center region. For the rectangular scattering
center region, $G_{2(4)}^{zc}$ is in general not equal to
$G_{2(4)}^{cz}$.
\bibitem{note5}
Here the parameters of the material "n-type HgTe" are used in the
numerical calculation, which is the same as the Ref.\cite{Gui}. In
fact, for other materials (e.g. GaAs with $m^* =0.036 m_e$),
the numerical results are similar.
\end{thebibliography}
\end{document}